\documentclass[11pt]{article}
\usepackage{graphicx}
\usepackage{geometry}
\usepackage{multirow}
\usepackage{url}
\usepackage[square, sort&compress, numbers]{natbib}
\usepackage{hyperref}
\usepackage[utf8]{inputenc}

\newcommand{\be}{\begin{equation}}
\newcommand{\ee}{\end{equation}}
\newcommand{\msbar}{\overline{MS}}
\newcommand{\mll}{m_{\ell \ell}}
\newcommand{\mllhat}{\ensuremath{\hat{\mathrm{m}}_{\ell\ell}}}
\newcommand{\diff}{\ensuremath{\mathrm{d}}}
\newcommand{\invfb}{\ensuremath{\mathrm{fb}^{-1}}}

\def\Title#1{\begin{center} {\Large #1 } \end{center}}
\def\Author#1{\begin{center}{ \sc #1} \end{center}}
\def\Address#1{\begin{center}{ \small \it #1} \end{center}}

\newenvironment{Abstract}{\begin{quotation}  }{\end{quotation}}
\newenvironment{Presented}{\begin{quotation} \begin{center} 
             PRESENTED AT\end{center}\bigskip 
      \begin{center}\begin{large}}{\end{large}\end{center} \end{quotation}}

\begin{document}
\begin{titlepage}
\vfill
\Title{Probing the weak mixing angle at high energy}
\vfill
\Author{ Clara Lavinia Del Pio$^{a,b,*}$, Simone Amoroso$^c$, Mauro Chiesa$^b$, Ekaterina Lipka$^{c,d}$, Fulvio Piccinini$^b$, Federico Vazzoler$^c$, Alessandro Vicini$^{e,f}$}
\Address{(*) Speaker \\
(a) Dipartimento di Fisica, Universit\`a di Pavia, via A. Bassi 6, Pavia, Italy\\
(b) INFN, Sezione di Pavia, via A. Bassi
6, Pavia, Italy \\
(c) Deutsches Elektronen-Synchrotron DESY, Notkestr. 85, 22607 Hamburg, Germany \\
(d) Bergische Universit\"at Wuppertal, Gaußstrassse 20, Wuppertal, Germany \\
(e) Dipartimento di Fisica, Universit\`a degli Studi di Milano, via G. Celoria 16, Milano, Italy \\
(f) INFN, Sezione di Milano, via G. Celoria 16, Milano, Italy}
\vfill
\begin{Abstract}
The weak mixing angle is a probe of the vector-axial coupling structure of electroweak interactions. It has been measured precisely at the $Z$-pole by experiments at the LEP and SLD colliders, but its energy dependence above $M_Z$ remains unconstrained.

In this contribution we propose to exploit measurements of Neutral-Current Drell Yan at large invariant dilepton masses at the Large Hadron Collider, to determine the scale dependence of the weak mixing angle in the $\overline{MS}$ renormalisation scheme, $\sin^2 \theta_w^{\msbar}(\mu)$.
Such a measurement can be used to test the Standard Model predictions for the $\overline{MS}$ running at TeV scales, and to set model-independent constraints on new states with electroweak quantum numbers.
To this end, we present an implementation of $\sin^2 \theta_w^{\msbar}(\mu)$ in the \textsc{POWHEG-BOX} Monte Carlo event generator, which we use to explore the potential of future analyses with the LHC Run~3 and High-Luminosity datasets. In particular, the impact of the higher order corrections and of the uncertainties due to the knowledge of parton distribution functions are studied. This contribution is based on~\cite{amoroso2023probing}.
\end{Abstract}
\vspace{0.3cm}
\begin{Presented}
DIS2023: XXX International Workshop on Deep-Inelastic Scattering and
Related Subjects, \\
Michigan State University, USA, 27-31 March 2023 \\
     \includegraphics[width=9cm]{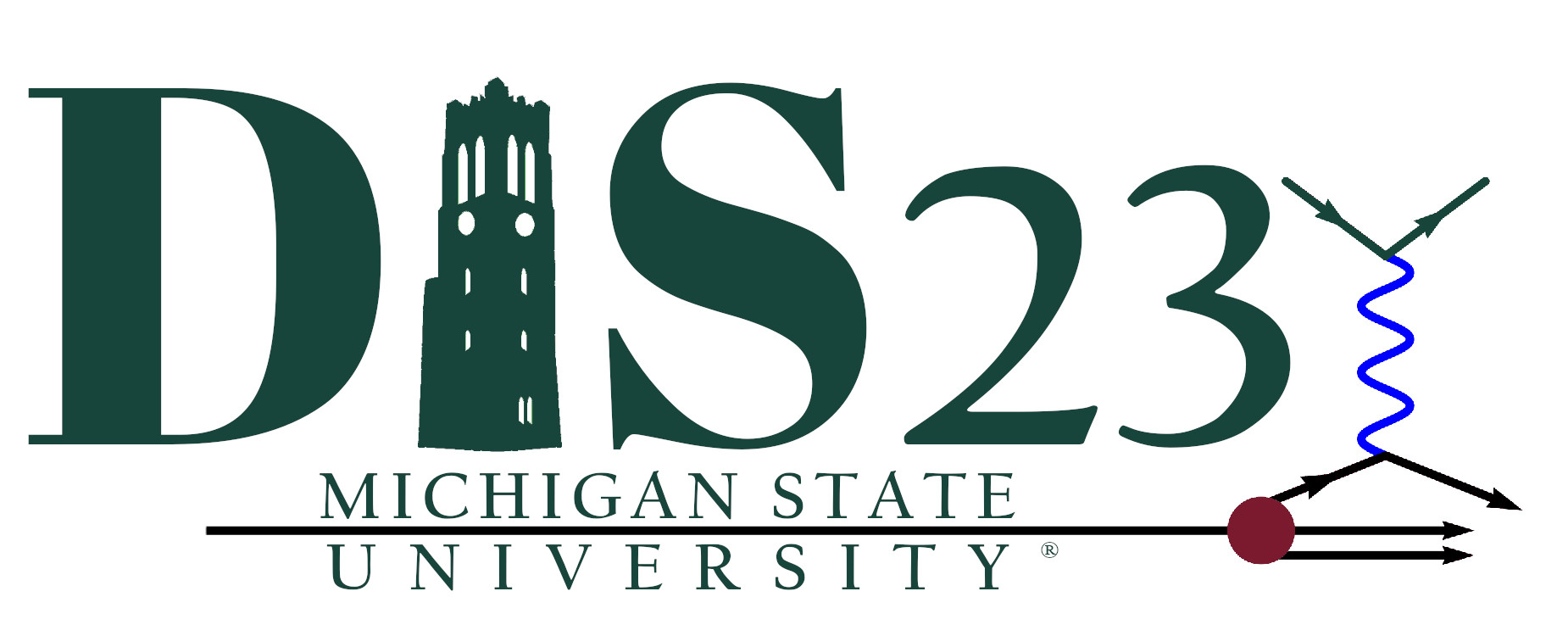}
\end{Presented}
\vfill
\end{titlepage}

\section{Introduction}
\label{section:introduction}
The electroweak mixing angle, $\theta_w$, is one of the fundamental parameters of the Standard Model (SM) of particle physics. It has a crucial role in the gauge structure of the electroweak interaction, as it regulates the mixing of the unphysical fields to give the photon and $Z$-boson fields, and it enters the unification relation of the weak and electromagnetic forces.
In the electroweak (EW) SM Lagrangian it is defined as:
\begin{equation}
    \label{eq:lagrangian-definition}
    \sin \theta_w= \frac{e}{g_2} = \frac{g_1}{\sqrt{g_1^2+g_2^2}},
\end{equation}
where $e$ is the positron charge, while $g_1$ and $g_2$ are the $U(1)_Y$ and $SU(2)_L$ gauge couplings.
In the on-shell renormalization scheme, the weak mixing angle is defined in terms of the $W$ and $Z$ masses, in such a way that the relation:
\be
 \sin^2 \theta_w^{OS} = 1-\frac{M_W^2}{M_Z^2}
\ee
holds at all orders in perturbation theory. Another possible definition is the effective one, that is introduced at the $Z$-boson peak, namely:
\be
\sin^2 \theta^f_{eff} \, = \, \frac{1}{4 |Q^f|} \left( 1 - {\rm Re} \frac{g_V^f}{g_A^f} \right) \, ,
\ee
where $Q_f$ is the electric charge of fermion $f$ in units of the positron charge, and $g_{V(A)}^f$ are the effective vector (axial-vector) couplings of fermion $f$ to the $Z$ boson.
At tree-level, all definitions coincide, but they start to differ
when considering radiative corrections, acquiring a dependence on the renormalization scheme and on the input parameter scheme used.

In the modified minimal-subtraction ($\overline{MS}$) renormalization scheme, the one used here, the running quantity $\sin^2 \theta_w^{\msbar}(\mu)$ is defined as:
\begin{equation}
    \label{eq:running-couplings}
    \sin^2\theta_{w}^{\msbar}(\mu) \equiv \frac{4\pi \alpha^{\msbar}(\mu)}{{g_2^2}^{\msbar}(\mu)},
\end{equation}
where $\mu$ is the renormalization scale and $\alpha^{\msbar}(\mu)$ is the running electromagnetic coupling.

The EW mixing angle at the $M_Z$ scale has been determined by using the effective definition $\sin^2 \theta^f_{eff}$, at both leptonic~\cite{ALEPH:2005ab}
and hadronic colliders~\cite{CDF:2018cnj,ATLAS:2015ihy,CMS:2018ktx,LHCb:2015jyu}, 
with a precision at the sub-percent level. At low energies $\sin^2\theta_{w}$ has been extracted via measurements of atomic parity violation, neutrino, and polarised electron scattering on fixed targets~\cite{Kumar:2013yoa,Wood:1997,PhysRevLett.82.2484,Gu_na_2005,Antypas_2018,PhysRevLett.88.091802,Anthony_2005,Qweak_2018,PVDIS:2014cmd}.

Although some results on the EW mixing angle at large space-like scales have been obtained from deep inelastic scattering (DIS) data~\cite{ZEUS:2016vyd,H1:2018mkk}, the running at time-like scales above the $Z$-boson mass has never been experimentally probed.  Measuring the running of $\sin^2\theta_{w}$ at high energies serves as an important test of the SM consistency, while, on the other hand, the high energy regime offers an indirect means for probing new states carrying EW quantum numbers, potentially leading to modifications in the running of the EW gauge couplings\cite{Georgi:1974yf,Einhorn:1981sx}.
The question this contribution tries to answer is: will it be possible to test the running of $\sin^2\theta_{w}^{\msbar}(\mu)$ at high energies at the LHC?

\section{Implementation of the running in POWHEG-BOX}
The sensitivity to $\sin^2\theta_{w}$ is here studied by exploiting the substantial dataset of neutral-current Drell-Yan (NCDY) events at high dilepton invariant masses ($\mll$) expected to be produced in proton-proton collisions at the LHC, with a center-of-mass energy of $\sqrt{s}=13.6$~TeV.

To improve the accuracy of existing analyses, that rely on leading order (LO) EW matrix elements, a full EW next-to-leading order (NLO) calculation has been implemented. It features a hybrid renormalization scheme, where the Lagrangian parameters $e$ and $\sin^2\theta_{w}$ are renormalized in the $\msbar$ scheme, by following the convention in~\cite{Erler:2004in}, while the $Z$-boson mass is taken in the on-shell scheme: $(\alpha^{\msbar}(\mu),\sin^2\theta_{w}^{\msbar}(\mu),M_Z)$. The code lies in the framework of an upgraded version~\cite{Chiesa:in-prep}
 of the $\textsc{Z\_ew-BMNNPV}$ package~\cite{Barze:2013fru} 
 of the \textsc{POWHEG-BOX}~\cite{Nason:2004rx,Frixione:2007vw,Alioli:2010xd} Monte Carlo (MC) event generator. 
In the presented study, the running parameters are calculated by activating the decoupling option for the $W$ boson and top quark for $\mu < M_W$ and $\mu < m_{\rm top}$, respectively, while we set to zero the $\mathcal{O}(\alpha)$ threshold corrections.

\section{Analysis and fit strategy}
We investigate the triple differential NCDY cross sections as a function of the dilepton invariant mass $\mll$, rapidity $y_{\ell \ell}$, and of the cosine of the angle between the incoming and outgoing fermions in the Collins-Soper reference frame, $\theta_{CS}$:
\be
\frac{\diff^3 \sigma}{\diff \mll  \, \diff y_{\ell \ell} \,  \diff\cos\theta_{CS}} \, .
\ee
At the $Z$ peak, the EW mixing angle has been usually extracted by measuring the forward-backward asymmetry $A_{FB}$, but
at high energy the absolute differential cross section is a more suitable observable for the extraction of $\sin^2\theta_{w}^{\msbar}(\mu)$, as it can be seen by defining the sensitivity to $\sin^2\theta_{w}^{\msbar}(\mu)$ as the logarithmic derivative multiplied by $\sin^2\theta_{w}^{\msbar}(\mu)$, that at $1$~TeV is found to be three times larger for the cross-sections than for $A_{FB}$.

We consider NCDY production at the LHC by analysing two scenarios: LHC Run~3 (integrated luminositiy of $300$ $\invfb$) and High-Luminosity LHC (HL-LHC, $3000$ $\invfb$).
We evaluate the triple differential NCDY cross section in six bins in $\mll$, from $116$~GeV up to $5000$~GeV, six bins in $|y_{\ell \ell}|$ from $0.0$ to $2.5$, and two bins in $\cos \theta_{CS}$ for the forward and backward directions.
Fiducial selections, usually employed in ATLAS and CMS measurements, are applied to the leptons, that are defined at Born level.

The MC predictions are generated at NLO QCD+NLO EW with \textsc{POWHEG-BOX}, excluding photonic corrections, and are then interfaced to \textsc{PYTHIA8.307}~\cite{Sjostrand:2014zea} to include the effect of parton showering, underlying event, hadronization and QED radiation from quarks. 

Since we are using a scheme with $\sin^2\theta_{w}^{\msbar}(\mu)$ in input, we can consistently generate templates at any order in perturbation theory, by varying the $\msbar$ parameter to be determined in the template fit procedure, thus allowing a solid precision determination of $\sin^2\theta_{w}^{\msbar}(\mu)$ at hadronic colliders.
The input EW parameters for the generation of the nominal pseudo-data are set to their $\msbar$ values taken at $\mu = M_Z$. Templates are generated assuming the SM running of $\alpha^{\msbar}(\mu)$, and setting the initial condition for the running of $\sin^2\theta_{w}^{\msbar}(\mu)$ in each $\mll$ bin to the expected SM prediction at the central value of the bin $\hat{m}_{\ell \ell}$, varied by $\pm0.01$. 
In each $\mll$ bin, $10^9$ MC events are generated. 

Detector response is emulated with parameterized lepton efficiencies and resolutions, inspired by those derived by ATLAS at Run~2~\cite{ATLAS:2019jvq,ATLAS:2015lne,ATLAS:2022jjr} and evaluated with \textsc{RIVET}~\cite{Buckley:2010ar}.

\section{Uncertainties}

Statistical uncertainties in the pseudo-data are computed from the predicted number of events at reconstructed level in each bin. 
Systematic uncertainties in the lepton reconstruction and efficiencies are taken by ATLAS measurements at Run~2~\cite{ATLAS:2016gic} and extrapolated to the working conditions at Run~3 and HL-LHC, assuming a reduction factor of $2$ and $4$, respectively, while the uncertainty in the luminosity determination is $1.5\%$ for Run~3 and $1\%$ for HL-LHC.

We estimate theoretical uncertainties due to PDFs by propagating the \\ \verb|NNPDF31_nnlo_as_0118_hessian| eigenvectors, and PDF variations by using grids generated with \verb|Madgraph_aMC@NLO| and \verb|aMCfast|~\cite{Bertone:2014zva,Alwall:2014hca}.
Since the NCDY production cross section is known up to N3LO in $\alpha_S$, the \verb|n3loxs| code~\cite{Baglio:2022wzu}
is employed to compute cross-sections and 7-point variations of the renormalization and factorization scales, $\mu_R$ and $\mu_F$, as a function of $\mll$ at N3LO in QCD, finding that N3LO corrections to the cross section are small, $2\%$ at maximum.
The uncertainty associated to missing EW higher-order contributions is estimated by squaring the size of the NLO weak correction, that amounts to a maximum of $1\%$ in the last $\mll$ bin considered.
Within the $\msbar$ scheme, an alternative way to quantify this source of uncertainty is given by a 2-point scale variation in $\mu$, which results in a change of the cross sections by about $0.1\%$ at NLO w.r.t. some $\%$ at LO.

\begin{figure}[t]
    \centering
    \includegraphics[width = 0.7\textwidth]{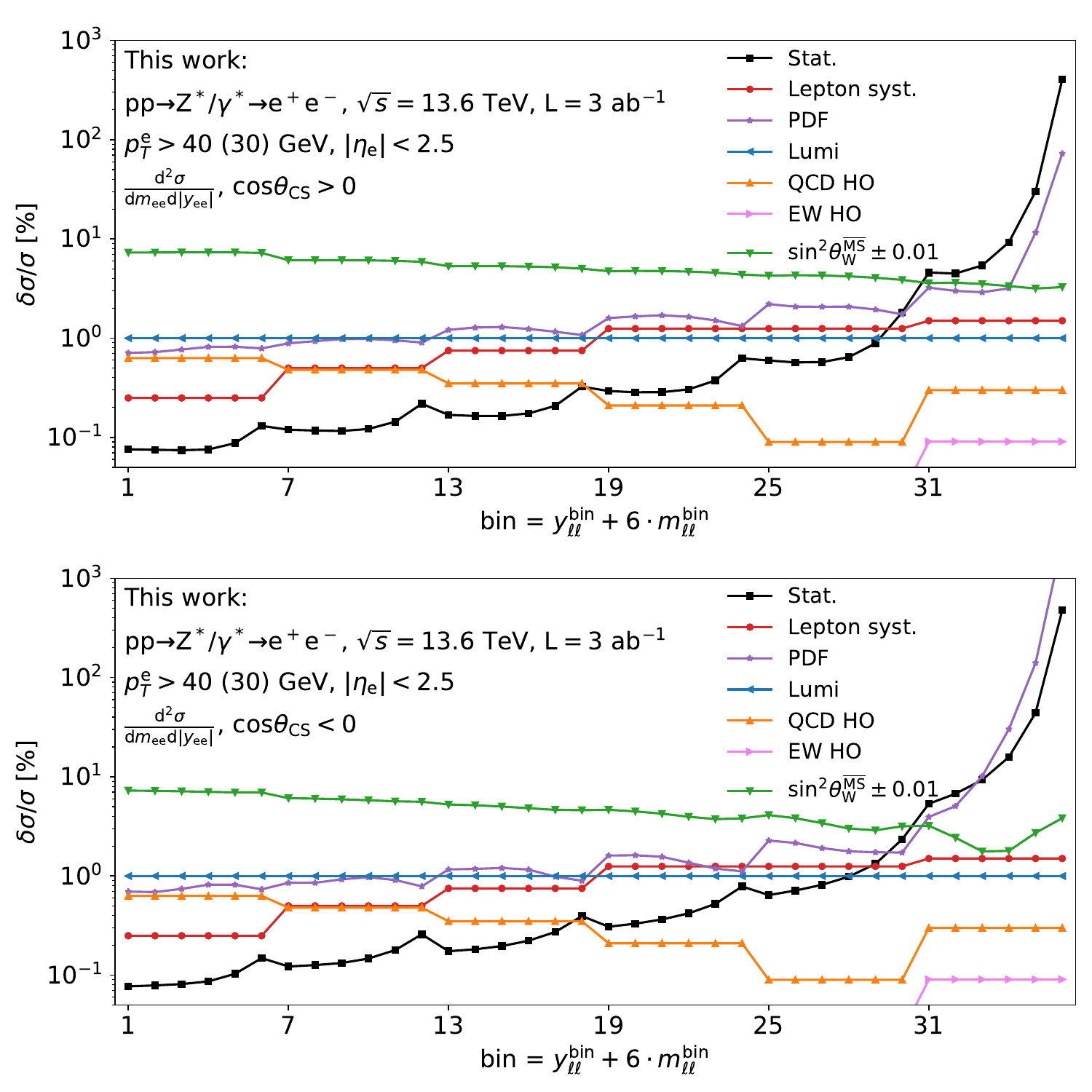}
    \caption{\footnotesize Relative contribution of the different sources of uncertainty to the triple differential cross section $\diff\sigma/\diff|y_{\ell \ell}| \diff\mll$ in the forward (up) and backward (bottom) directions, for the electron channel in the HL-LHC scenario. The variation of $\sin^2\theta_{w}^{\msbar}(\mu=\hat{m}_{\ell \ell})$ by a factor $\pm0.01$ in each bin is also shown.}
    \label{fig:relative impact prefit}
\end{figure}

Fig.~\ref{fig:relative impact prefit} shows the contributions of different uncertainty sources to the triple differential NCDY cross sections in the electron channel for the HL-LHC scenario, together with the representative variation of $\sin^2\theta_{w}^{\msbar}(\mu=\hat{m}_{\ell \ell})$ by $\pm0.01$. 
Results in the muon channel are similar.

\section{Results}
\begin{figure}[t]
    \centering
    \includegraphics[width = 0.7\textwidth]{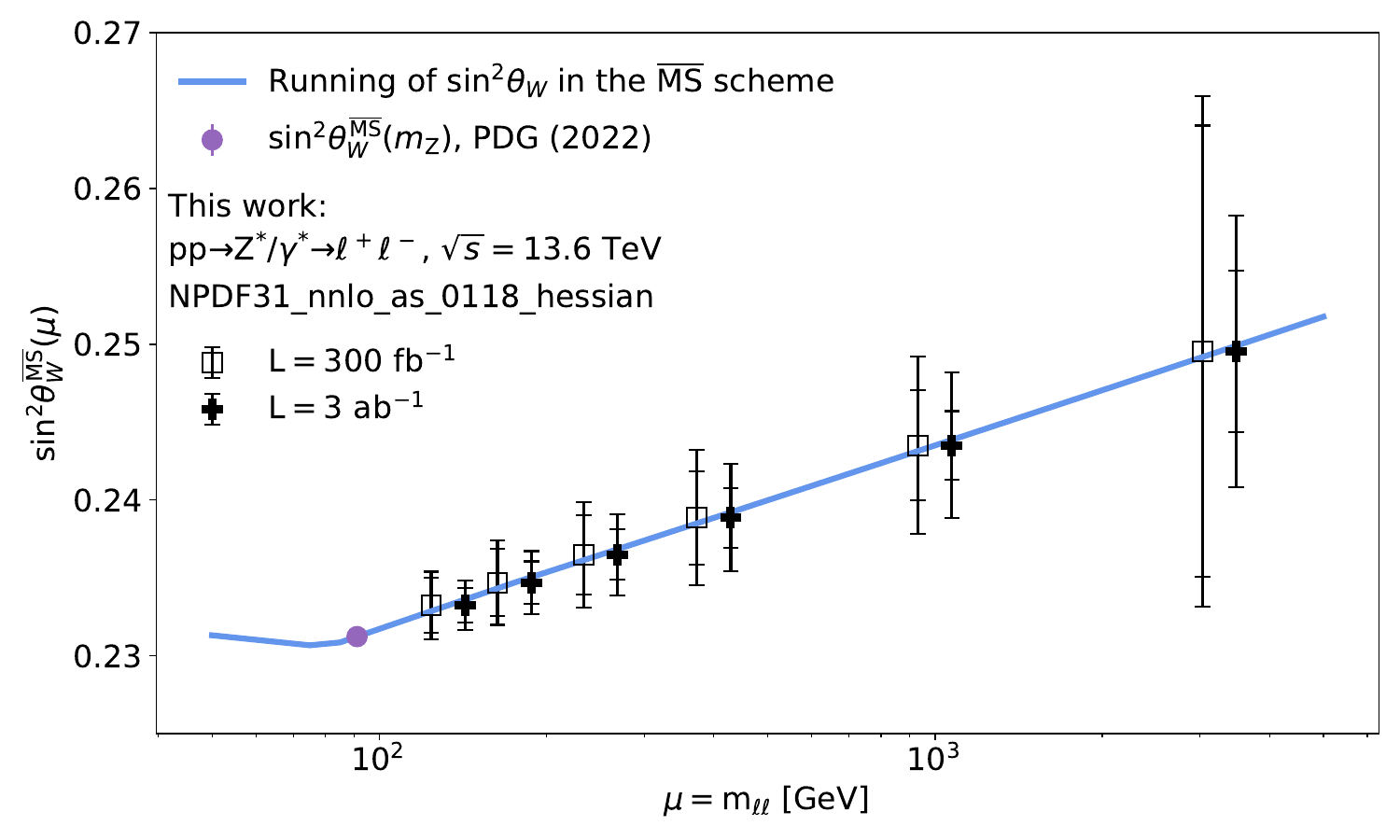}
    \caption{\footnotesize SM running of the EW mixing angle (blue line), together with the combined experimental measurement at $\mu=M_Z$ (violet point).
    The expected results obtained in our analysis are shown in black crosses (black squares) for the LHC Run~3 (HL-LHC), shifted for readibility to left and right, respectively.
    The outer error bars are the total expected uncertainty on $\sin^2\theta_{w}^{\msbar}(\mu)$, while the inner error bars include only statistical and experimental uncertainties.
    \label{fig:results for the scenarios}}
\end{figure}

The sensitivity to the running is assessed by extracting the expected value of $\sin^2\theta_{w}^{\msbar}(\mu)$ and evaluating its uncertainty $\delta\sin^2\theta_{w}^{\msbar}(\mu)$ as a function of $\mllhat$, assuming SM running for $\alpha^{\msbar}(\mu)$. 
The expected $\delta\sin^2\theta_{w}^{\msbar}(\mu)$ values for each $\mll$ bin are simultaneously obtained in the fit to the triple differential cross section pseudo-data, by minimising a $\chi^2$ function with the xFitter analysis tool~\cite{Alekhin:2014irh}. 
When computing the $\chi^2$, the cross-section is considered linearly dependent on $\delta \sin^2\theta_{w}^{\msbar}(\mu)$, while
the expected statistical and experimental systematic uncertainties, the theoretical uncertainties from PDFs and missing higher orders are included as nuisance parameters and are constrained in the fit.
The obtained values of $\delta\sin^2\theta_{w}^{\msbar}(\mu)$ are presented in Fig.~\ref{fig:results for the scenarios}. They range from about $1\%$ ($1\%$) to $7\%$ ($3\%$) for the LHC Run~3 (HL-LHC) scenario.

The largest contribution to the uncertainty on $\delta \sin^2\theta_{W}^\mathrm{\overline{MS}}(\mu)$ comes from the PDFs at large $x$.
We repeat the fit using alternative PDF sets, one of them including also the photon, and find that
the contribution of the PDF uncertainty to $\delta \sin^2\theta_{W}^\mathrm{\overline{MS}}(\mu)$ amounts to some $\%$ at maximum and can vary significantly with the PDF set used, by up to a factor $2$ in the last $\mll$ bin. Future PDFs fits and analyses could however improve these figures by the time HL-LHC will start to run.

\section{Conclusions and future prospects}
In this contribution, we show that NCDY measurements at current and future LHC runs will probe the $\msbar$ running of the electroweak mixing angle with a precision at the percent level, by using the triple differential NCDY cross sections in $\mll$, $y_{\ell \ell}$ and $\cos \theta_{CS}$  and under the assumption of SM running of the electromagnetic coupling constant. 

A future determination of $\sin^2\theta_{W}^\mathrm{\overline{MS}}(\mu)$ at the TeV scale will allow a consistency test of the SM, as it can be connected to low-energy existing measurements via the evolution predicted by the renormalization group equation. The present analysis is framed in a general context where several measurements are foreseen to probe the weak mixing angle with high precision at the $Z$ pole and higher energies~\cite{Mangano:2651294,Abelleira_Fernandez_2012,baer2013international,linssen2012physics,thecepcstudygroup2018cepc}, as well as at lower scales~\cite[\& references therein]{Zhang:2016czr,Versolato:2010,PhysRevA.81.032114,mollercollaboration2014,Becker_2018,Souder:2012zz,Boughezal_2022,abraham2023neutrino}.

Moreover, a high energy determination of the weak mixing angle could indirectly probe new physics particles, provided that they carry EW quantum numbers and thus enter the running of the EW gauge couplings. The pure SM approach presented here is particularly valuable as it is complementary to general methods that make use of Effective Field Theories~\cite{DiLuzio:2018jwd,Torre:2020aiz,Alioli:2020kez}, and could be an additional useful tool to explore new physics beyond the SM.

\footnotesize
\bibliography{biblio.bib} 

\providecommand{\href}[2]{#2}\begingroup\raggedright\begin{thebibliography}{10}

\bibitem{amoroso2023probing}
S.~Amoroso, M.~Chiesa, C.~L.~D. Pio, K.~Lipka, F.~Piccinini, F.~Vazzoler, and
  A.~Vicini, ``{Probing the weak mixing angle at high energies at the LHC and
  HL-LHC}.'' 2023.
\newblock \href{https://arxiv.org/abs/2302.10782}{{\ttfamily arXiv:2302.10782
  [hep-ph]}}.

\bibitem{ALEPH:2005ab}
{ ALEPH, DELPHI, L3, OPAL, SLD, LEP Electroweak Working Group, SLD Electroweak
  Group, SLD Heavy Flavour Group} Collaboration, {ALEPH, DELPHI, L3, OPAL, SLD,
  LEP Electroweak Working Group, SLD Electroweak Group, SLD Heavy Flavour
  Group}, ``{Precision electroweak measurements on the $Z$ resonance},''
  \href{https://dx.doi.org/10.1016/j.physrep.2005.12.006}{{\em Phys. Rept.}
  {\bfseries 427} (2006) 257--454},
  \href{https://arxiv.org/abs/hep-ex/0509008}{{\ttfamily
  arXiv:hep-ex/0509008}}.

\bibitem{CDF:2018cnj}
{ CDF, D0} Collaboration, {CDF and D0 Collaborations}, ``{Tevatron Run II
  combination of the effective leptonic electroweak mixing angle},''
  \href{https://dx.doi.org/10.1103/PhysRevD.97.112007}{{\em Phys. Rev. D}
  {\bfseries 97} no.~11, (2018) 112007},
  \href{https://arxiv.org/abs/1801.06283}{{\ttfamily arXiv:1801.06283
  [hep-ex]}}.

\bibitem{ATLAS:2015ihy}
{ ATLAS} Collaboration, {ATLAS Collaboration}, ``{Measurement of the
  forward-backward asymmetry of electron and muon pair-production in $pp$
  collisions at $\sqrt{s}$ = 7 TeV with the ATLAS detector},''
  \href{https://dx.doi.org/10.1007/JHEP09(2015)049}{{\em JHEP} {\bfseries 09}
  (2015) 049}, \href{https://arxiv.org/abs/1503.03709}{{\ttfamily
  arXiv:1503.03709 [hep-ex]}}.

\bibitem{CMS:2018ktx}
{ CMS} Collaboration, {CMS Collaboration}, ``{Measurement of the weak mixing
  angle using the forward-backward asymmetry of Drell-Yan events in pp
  collisions at 8 TeV},''
  \href{https://dx.doi.org/10.1140/epjc/s10052-018-6148-7}{{\em Eur. Phys. J.
  C} {\bfseries 78} no.~9, (2018) 701},
  \href{https://arxiv.org/abs/1806.00863}{{\ttfamily arXiv:1806.00863
  [hep-ex]}}.

\bibitem{LHCb:2015jyu}
{ LHCb} Collaboration, {LHCb Collaboration}, ``{Measurement of the
  forward-backward asymmetry in $Z/\gamma^{\ast} \rightarrow \mu^{+}\mu^{-}$
  decays and determination of the effective weak mixing angle},''
  \href{https://dx.doi.org/10.1007/JHEP11(2015)190}{{\em JHEP} {\bfseries 11}
  (2015) 190}, \href{https://arxiv.org/abs/1509.07645}{{\ttfamily
  arXiv:1509.07645 [hep-ex]}}.

\bibitem{Kumar:2013yoa}
K.~S. Kumar, S.~Mantry, W.~J. Marciano, and P.~A. Souder, ``{Low Energy
  Measurements of the Weak Mixing Angle},''
  \href{https://dx.doi.org/10.1146/annurev-nucl-102212-170556}{{\em Ann. Rev.
  Nucl. Part. Sci.} {\bfseries 63} (2013) 237--267},
  \href{https://arxiv.org/abs/1302.6263}{{\ttfamily arXiv:1302.6263 [hep-ex]}}.

\bibitem{Wood:1997}
C.~Wood {\em et~al.}, ``Measurement of parity nonconservation and an anapole
  moment in cesium,''
  \href{https://dx.doi.org/10.1126/science.275.5307.1759}{{\em Science}
  {\bfseries 275} no.~5307, (1997) 1759--1763}.

\bibitem{PhysRevLett.82.2484}
S.~C. Bennett and C.~E. Wieman, ``Measurement of the
  $6\mathit{S}\ensuremath{\rightarrow}7\mathit{S}$ transition polarizability in
  atomic cesium and an improved test of the standard model,''
  \href{https://dx.doi.org/10.1103/PhysRevLett.82.2484}{{\em Phys. Rev. Lett.}
  {\bfseries 82} (1999) 2484--2487}.
  \url{https://link.aps.org/doi/10.1103/PhysRevLett.82.2484}.

\bibitem{Gu_na_2005}
J.~Gu{\'{e}}na, M.~Lintz, and M.~A. Bouchiat, ``Measurement of the parity
  violating $6s-7s$ transition amplitude in cesium achieved within $2 \times
  10^{-13}$ atomic-unit accuracy by stimulated-emission detection,''
  \href{https://dx.doi.org/10.1103/physreva.71.042108}{{\em Physical Review A}
  {\bfseries 71} no.~4, (2005) }.
  \url{https://doi.org/10.1103%2Fphysreva.71.042108}.

\bibitem{Antypas_2018}
D.~Antypas {\em et~al.}, ``Isotopic variation of parity violation in atomic
  ytterbium,'' \href{https://dx.doi.org/10.1038/s41567-018-0312-8}{{\em Nature
  Physics} {\bfseries 15} no.~2, (2018) 120--123}.
  \url{https://doi.org/10.1038%2Fs41567-018-0312-8}.

\bibitem{PhysRevLett.88.091802}
{NuTeV Collaboration}, ``Precise determination of electroweak parameters in
  neutrino-nucleon scattering,''
  \href{https://dx.doi.org/10.1103/PhysRevLett.88.091802}{{\em Phys. Rev.
  Lett.} {\bfseries 88} (2002) 091802}.
  \url{https://link.aps.org/doi/10.1103/PhysRevLett.88.091802}.

\bibitem{Anthony_2005}
{SLAC E158 Collaboration}, ``Precision measurement of the weak mixing angle in
  m{\o}ller scattering,''
  \href{https://dx.doi.org/10.1103/physrevlett.95.081601}{{\em Physical Review
  Letters} {\bfseries 95} no.~8, (2005) }.
  \url{https://doi.org/10.1103%2Fphysrevlett.95.081601}.

\bibitem{Qweak_2018}
{JLAB Qweak Collaboration}, ``Precision measurement of the weak charge of the
  proton,'' \href{https://dx.doi.org/10.1038/s41586-018-0096-0}{{\em Nature}
  {\bfseries 557} no.~7704, (2018) 207--211}.
  \url{https://doi.org/10.1038%2Fs41586-018-0096-0}.

\bibitem{PVDIS:2014cmd}
{ PVDIS} Collaboration, D.~Wang {\em et~al.}, ``{Measurement of parity
  violation in electron\textendash{}quark scattering},''
  \href{https://dx.doi.org/10.1038/nature12964}{{\em Nature} {\bfseries 506}
  no.~7486, (2014) 67--70}.

\bibitem{ZEUS:2016vyd}
{ ZEUS} Collaboration, {ZEUS Collaboration}, ``{Combined QCD and electroweak
  analysis of HERA data},''
  \href{https://dx.doi.org/10.1103/PhysRevD.93.092002}{{\em Phys. Rev. D}
  {\bfseries 93} no.~9, (2016) 092002},
  \href{https://arxiv.org/abs/1603.09628}{{\ttfamily arXiv:1603.09628
  [hep-ex]}}.

\bibitem{H1:2018mkk}
{H1 Collaboration}, ``{Determination of electroweak parameters in polarised
  deep-inelastic scattering at HERA},''
  \href{https://dx.doi.org/10.1140/epjc/s10052-018-6236-8}{{\em Eur. Phys. J.
  C} {\bfseries 78} no.~9, (2018) 777},
  \href{https://arxiv.org/abs/1806.01176}{{\ttfamily arXiv:1806.01176
  [hep-ex]}}.

\bibitem{Georgi:1974yf}
H.~Georgi, H.~R. Quinn, and S.~Weinberg, ``{Hierarchy of Interactions in
  Unified Gauge Theories},''
  \href{https://dx.doi.org/10.1103/PhysRevLett.33.451}{{\em Phys. Rev. Lett.}
  {\bfseries 33} (1974) 451--454}.

\bibitem{Einhorn:1981sx}
M.~B. Einhorn and D.~R.~T. Jones, ``{The Weak Mixing Angle and Unification Mass
  in Supersymmetric SU(5)},''
  \href{https://dx.doi.org/10.1016/0550-3213(82)90502-8}{{\em Nucl. Phys. B}
  {\bfseries 196} (1982) 475--488}.

\bibitem{Erler:2004in}
J.~Erler and M.~J. Ramsey-Musolf, ``{The weak mixing angle at low energies},''
  \href{https://dx.doi.org/10.1103/PhysRevD.72.073003}{{\em Phys. Rev. D}
  {\bfseries 72} (2005) 073003},
  \href{https://arxiv.org/abs/hep-ph/0409169}{{\ttfamily
  arXiv:hep-ph/0409169}}.

\bibitem{Chiesa:in-prep}
M.~Chiesa, C.~L. Del~Pio, and F.~Piccinini, ``{On electroweak corrections to
  neutral current Drell-Yan with the POWHEG BOX},'' {\em in preparation} .

\bibitem{Barze:2013fru}
L.~Barzé {\em et~al.}, ``{Neutral current Drell-Yan with combined QCD and
  electroweak corrections in the POWHEG BOX},''
  \href{https://dx.doi.org/10.1140/epjc/s10052-013-2474-y}{{\em Eur. Phys. J.
  C} {\bfseries 73} no.~6, (2013) 2474},
  \href{https://arxiv.org/abs/1302.4606}{{\ttfamily arXiv:1302.4606 [hep-ph]}}.

\bibitem{Nason:2004rx}
P.~Nason, ``{A New method for combining NLO QCD with shower Monte Carlo
  algorithms},'' \href{https://dx.doi.org/10.1088/1126-6708/2004/11/040}{{\em
  JHEP} {\bfseries 11} (2004) 040},
  \href{https://arxiv.org/abs/hep-ph/0409146}{{\ttfamily
  arXiv:hep-ph/0409146}}.

\bibitem{Frixione:2007vw}
S.~Frixione, P.~Nason, and C.~Oleari, ``{Matching NLO QCD computations with
  Parton Shower simulations: the POWHEG method},''
  \href{https://dx.doi.org/10.1088/1126-6708/2007/11/070}{{\em JHEP} {\bfseries
  11} (2007) 070}, \href{https://arxiv.org/abs/0709.2092}{{\ttfamily
  arXiv:0709.2092 [hep-ph]}}.

\bibitem{Alioli:2010xd}
S.~Alioli, P.~Nason, C.~Oleari, and E.~Re, ``{A general framework for
  implementing NLO calculations in shower Monte Carlo programs: the POWHEG
  BOX},'' \href{https://dx.doi.org/10.1007/JHEP06(2010)043}{{\em JHEP}
  {\bfseries 06} (2010) 043}, \href{https://arxiv.org/abs/1002.2581}{{\ttfamily
  arXiv:1002.2581 [hep-ph]}}.

\bibitem{Sjostrand:2014zea}
T.~Sj\"ostrand {\em et~al.}, ``{An introduction to PYTHIA 8.2}''
  \href{https://dx.doi.org/10.1016/j.cpc.2015.01.024}{{\em Comput. Phys.
  Commun.} {\bfseries 191} (2015) 159--177},
  \href{https://arxiv.org/abs/1410.3012}{{\ttfamily arXiv:1410.3012 [hep-ph]}}.

\bibitem{ATLAS:2019jvq}
{ ATLAS} Collaboration, {ATLAS Collaboration}, ``{Electron reconstruction and
  identification in the ATLAS experiment using the 2015 and 2016 LHC
  proton-proton collision data at $\sqrt{s}$ = 13 TeV},''
  \href{https://dx.doi.org/10.1140/epjc/s10052-019-7140-6}{{\em Eur. Phys. J.
  C} {\bfseries 79} no.~8, (2019) 639},
  \href{https://arxiv.org/abs/1902.04655}{{\ttfamily arXiv:1902.04655
  [physics.ins-det]}}.

\bibitem{ATLAS:2015lne}
{ ATLAS} Collaboration, {ATLAS Collaboration}, ``{Muon reconstruction
  performance in early $\sqrt{s}$ = 13 TeV data},''.
  \url{http://cds.cern.ch/record/2047831}.

\bibitem{ATLAS:2022jjr}
{ ATLAS} Collaboration, {ATLAS Collaboration}, ``{Studies of the muon momentum
  calibration and performance of the ATLAS detector with $pp$ collisions at
  $\sqrt{s}$=13 TeV},'' \href{https://arxiv.org/abs/2212.07338}{{\ttfamily
  arXiv:2212.07338 [hep-ex]}}.

\bibitem{Buckley:2010ar}
A.~Buckley {\em et~al.}, ``{Rivet user manual},''
  \href{https://dx.doi.org/10.1016/j.cpc.2013.05.021}{{\em Comput. Phys.
  Commun.} {\bfseries 184} (2013) 2803--2819},
  \href{https://arxiv.org/abs/1003.0694}{{\ttfamily arXiv:1003.0694 [hep-ph]}}.

\bibitem{ATLAS:2016gic}
{ ATLAS} Collaboration, {ATLAS Collaboration}, ``{Measurement of the
  double-differential high-mass Drell-Yan cross section in pp collisions at $
  \sqrt{s}=8 $ TeV with the ATLAS detector},''
  \href{https://dx.doi.org/10.1007/JHEP08(2016)009}{{\em JHEP} {\bfseries 08}
  (2016) 009}, \href{https://arxiv.org/abs/1606.01736}{{\ttfamily
  arXiv:1606.01736 [hep-ex]}}.

\bibitem{Bertone:2014zva}
V.~Bertone {\em et~al.}, ``{aMCfast: automation of fast NLO computations for
  PDF fits},'' \href{https://dx.doi.org/10.1007/JHEP08(2014)166}{{\em JHEP}
  {\bfseries 08} (2014) 166}, \href{https://arxiv.org/abs/1406.7693}{{\ttfamily
  arXiv:1406.7693 [hep-ph]}}.

\bibitem{Alwall:2014hca}
J.~Alwall {\em et~al.}, ``{The automated computation of tree-level and
  next-to-leading order differential cross sections, and their matching to
  parton shower simulations},''
  \href{https://dx.doi.org/10.1007/JHEP07(2014)079}{{\em JHEP} {\bfseries 07}
  (2014) 079}, \href{https://arxiv.org/abs/1405.0301}{{\ttfamily
  arXiv:1405.0301 [hep-ph]}}.

\bibitem{Baglio:2022wzu}
J.~Baglio, C.~Duhr, B.~Mistlberger, and R.~Szafron, ``{Inclusive production
  cross sections at N$^{3}$LO},''
  \href{https://dx.doi.org/10.1007/JHEP12(2022)066}{{\em JHEP} {\bfseries 12}
  (2022) 066}, \href{https://arxiv.org/abs/2209.06138}{{\ttfamily
  arXiv:2209.06138 [hep-ph]}}.

\bibitem{Alekhin:2014irh}
S.~Alekhin {\em et~al.}, ``{HERAFitter},''
  \href{https://dx.doi.org/10.1140/epjc/s10052-015-3480-z}{{\em Eur. Phys. J.
  C} {\bfseries 75} no.~7, (2015) 304},
  \href{https://arxiv.org/abs/1410.4412}{{\ttfamily arXiv:1410.4412 [hep-ph]}}.

\bibitem{Mangano:2651294}
M.~Mangano {\em et~al.},
  \href{https://dx.doi.org/10.1140/epjc/s10052-019-6904-3}{``{FCC Physics
  Opportunities: Future Circular Collider Conceptual Design Report Volume 1.
  Future Circular Collider},''} Tech. Rep.~6, CERN, Geneva, 2019.
\newblock \url{https://cds.cern.ch/record/2651294}.

\bibitem{Abelleira_Fernandez_2012}
J.~L.~A. Fernandez {\em et~al.}, ``A large hadron electron collider at {CERN}
  report on the physics and design concepts for machine and detector,''
  \href{https://dx.doi.org/10.1088/0954-3899/39/7/075001}{{\em Journal of
  Physics G: Nuclear and Particle Physics} {\bfseries 39} no.~7, (2012)
  075001}. \url{https://doi.org/10.1088%2F0954-3899%2F39%2F7%2F075001}.

\bibitem{baer2013international}
H.~Baer {\em et~al.}, ``{The {I}nternational {L}inear {C}ollider Technical
  Design Report - Volume 2: Physics}.'' 2013.
\newblock \href{https://arxiv.org/abs/1306.6352}{{\ttfamily arXiv:1306.6352
  [hep-ph]}}.

\bibitem{linssen2012physics}
L.~Linssen, A.~Miyamoto, M.~Stanitzki, and H.~Weerts, ``{Physics and Detectors
  at CLIC: CLIC Conceptual Design Report}.'' 2012.
\newblock \href{https://arxiv.org/abs/1202.5940}{{\ttfamily arXiv:1202.5940
  [physics.ins-det]}}.

\bibitem{thecepcstudygroup2018cepc}
T.~C.~S. Group, ``{CEPC Conceptual Design Report: Volume 2 - Physics $\&$
  Detector}.'' 2018.
\newblock \href{https://arxiv.org/abs/1811.10545}{{\ttfamily arXiv:1811.10545
  [hep-ex]}}.

\bibitem{Zhang:2016czr}
J.~Zhang {\em et~al.}, ``{Efficient inter-trap transfer of cold francium
  atoms},'' \href{https://dx.doi.org/10.1007/s10751-016-1347-9}{{\em Hyperfine
  Interact.} {\bfseries 237} no.~1, (2016) 150}.

\bibitem{Versolato:2010}
O.~Versolato {\em et~al.}, ``Laser spectroscopy of trapped short-lived ra+
  ions,'' \href{https://dx.doi.org/10.1103/PhysRevA.82.010501}{{\em Physical
  Review A} {\bfseries 82} no.~1, (2010) }.

\bibitem{PhysRevA.81.032114}
K.~Tsigutkin, D.~Dounas-Frazer, A.~Family, J.~E. Stalnaker, V.~V. Yashchuk, and
  D.~Budker, ``Parity violation in atomic ytterbium: Experimental sensitivity
  and systematics,'' \href{https://dx.doi.org/10.1103/PhysRevA.81.032114}{{\em
  Phys. Rev. A} {\bfseries 81} (Mar, 2010) 032114}.
  \url{https://link.aps.org/doi/10.1103/PhysRevA.81.032114}.

\bibitem{mollercollaboration2014}
{MOLLER Collaboration}, ``{The MOLLER Experiment: An Ultra-Precise Measurement
  of the Weak Mixing Angle Using M{\o}ller Scattering}.'' 2014.
\newblock \href{https://arxiv.org/abs/1411.4088}{{\ttfamily arXiv:1411.4088
  [nucl-ex]}}.

\bibitem{Becker_2018}
D.~Becker {\em et~al.}, ``{The P2 experiment},''
  \href{https://dx.doi.org/10.1140/epja/i2018-12611-6}{{\em The European
  Physical Journal A} {\bfseries 54} no.~11, (2018) }.
  \url{https://doi.org/10.1140%2Fepja%2Fi2018-12611-6}.

\bibitem{Souder:2012zz}
{ H1, ZEUS} Collaboration, P.~A. Souder, ``{Parity-violating PVDIS with
  SoLID},'' \href{https://dx.doi.org/10.1063/1.3700490}{{\em AIP Conf. Proc.}
  {\bfseries 1441} no.~1, (2012) 123--125}.

\bibitem{Boughezal_2022}
R.~Boughezal {\em et~al.}, ``Neutral-current electroweak physics and {SMEFT}
  studies at the {EIC},''
  \href{https://dx.doi.org/10.1103/physrevd.106.016006}{{\em Phys. Rev. D}
  {\bfseries 106} no.~1, (2022) }.
  \url{https://doi.org/10.1103%2Fphysrevd.106.016006}.

\bibitem{abraham2023neutrino}
R.~M. Abraham, S.~Foroughi-Abari, F.~Kling, and Y.-D. Tsai, ``{Neutrino
  Electromagnetic Properties and the Weak Mixing Angle at the LHC Forward
  Physics Facility}.'' 2023.
\newblock \href{https://arxiv.org/abs/2301.10254}{{\ttfamily arXiv:2301.10254
  [hep-ph]}}.

\bibitem{DiLuzio:2018jwd}
L.~Di~Luzio, R.~Gr\"ober, and G.~Panico, ``{Probing new electroweak states via
  precision measurements at the LHC and future colliders},''
  \href{https://dx.doi.org/10.1007/JHEP01(2019)011}{{\em JHEP} {\bfseries 01}
  (2019) 011}, \href{https://arxiv.org/abs/1810.10993}{{\ttfamily
  arXiv:1810.10993 [hep-ph]}}.

\bibitem{Torre:2020aiz}
R.~Torre, L.~Ricci, and A.~Wulzer, ``{On the W\&Y interpretation of high-energy
  Drell-Yan measurements},''
  \href{https://dx.doi.org/10.1007/JHEP02(2021)144}{{\em JHEP} {\bfseries 02}
  (2021) 144}, \href{https://arxiv.org/abs/2008.12978}{{\ttfamily
  arXiv:2008.12978 [hep-ph]}}.

\bibitem{Alioli:2020kez}
S.~Alioli, R.~Boughezal, E.~Mereghetti, and F.~Petriello, ``{Novel angular
  dependence in Drell-Yan lepton production via dimension-8 operators},''
  \href{https://dx.doi.org/10.1016/j.physletb.2020.135703}{{\em Phys. Lett. B}
  {\bfseries 809} (2020) 135703},
  \href{https://arxiv.org/abs/2003.11615}{{\ttfamily arXiv:2003.11615
  [hep-ph]}}.

\end{thebibliography}\endgroup
\bibliographystyle{caps}

\end{document}